\documentclass[prb,twocolumn,preprintnumbers,amsmath,amssymb,superscriptaddress]{revtex4}
\usepackage{graphicx}% Include figure files
\usepackage{epsfig}
\usepackage{dcolumn} %decimal point spacing table
\usepackage{bm} %bold math mode

\begin{document}

\title{Observation of 300 K High Energy MagnetoDielectric Response in the Bilayer Manganite
(La$_{0.4}$Pr$_{0.6}$)$_{1.2}$Sr$_{1.8}$Mn$_2$O$_7$}

\author{Jinbo Cao}
\affiliation{Department of Chemistry, University of Tennessee,
Knoxville, TN 37996 USA}
\author{Ram C. Rai}
\affiliation{Department of Chemistry, University of Tennessee,
Knoxville, TN 37996 USA}
\author{Sonal Brown}
\affiliation{Department of Chemistry, University of Tennessee,
Knoxville, TN 37996 USA}
\author{Janice L. Musfeldt}
\affiliation{Department of Chemistry, University of Tennessee,
Knoxville, TN 37996 USA}
\author{Ronald Tackett}
\affiliation{Department of Physics, Wayne State University, Detroit,
MI 48201 USA}
\author{Gavin Lawes}
\affiliation{Department of Physics, Wayne State University, Detroit,
MI 48201 USA}
\author{Yongjie Wang}
\affiliation{National High Magnetic Field Laboratory, Florida State
University, Tallahassee, FL 32310 USA}
\author{Xing Wei}
\affiliation{National High Magnetic Field Laboratory, Florida State
University, Tallahassee, FL 32310 USA}
\author{Mircea Apostu}
\affiliation{Laboratoire de Physico-Chimie de l'Etat Solide, Univ.
 Paris-Sud., UMR 8182, 91405 Orsay, France}
\author{Ramanathan Suryanarayanan}
\affiliation{Laboratoire de Physico-Chimie de l'Etat Solide, Univ.
 Paris-Sud., UMR 8182, 91405 Orsay, France}
\author{Alexandre Revcolevschi}
\affiliation{Laboratoire de Physico-Chimie de l'Etat Solide, Univ.
 Paris-Sud., UMR 8182, 91405 Orsay, France}

\date{\today}

\maketitle \clearpage

The discovery of magnetoelectric coupling and in particular of the
magnetodielectric (MD) effect, e.g., field-induced change in
dielectric constant $\epsilon_1$, has attracted  attention due to
both potential utilization in multifunctional devices and  the
desire to understand the fundamental physics underlying multiferroic
materials. \cite{Hemberger2005, Lunkenheimer2005, Weber2006,
Kimura2003, Rogado2005, Subramanian2006, Hur2004, Lorenz2004} For
instance,  rare earth manganites such as TbMnO$_3$, HoMnO$_3$,
YMnO$_3$, and DyMn$_2$O$_5$ display sizable static  MD effects at
low temperature, demonstrating that dielectric contrast can be
achieved by physical tuning through various magnetic transitions in
$H-T$ space.\cite{Kimura2003, Hur2004, Lorenz2004} Cubic spinels
such as CdCr$_2$S$_4$ and HgCr$_2$S$_4$ present even larger static
MD contrast,  attributed to relaxational dynamics driven by
magnetization in the field. \cite{Hemberger2005, Lunkenheimer2005,
Weber2006} More recently,  300 K MD behavior was reported in mixed
valent LuFe$_2$O$_4$ and attributed to charge ordering
effects.\cite{Subramanian2006} Mn-doped BiFeO$_3$ also exhibits a
potentially tunable  room temperature MD effect due to enhanced
thermal fluctuations near the N\'eel temperature. \cite{Yang2005}
Discovery of large MD effects in other materials combined with
mechanistic understanding of magnetoelectric cross-coupling is of
fundamental interest and may hold promise for next-generation memory
devices.

Static dielectric properties are typically measured in a parallel
plate/capacitance geometry at low frequencies (Hz - MHz).
\cite{Subramanian2006, Weber2006, Hemberger2005} Challenges with
this technique include contact problems, potential dead layers, and
edge effects. \cite{Zhou1997, Stengel2006} A contactless technique,
such as optical spectroscopy, eliminates these issues. At the same
time, the electromagnetic spectrum is very broad. This opens the
possibility of exploiting changes in $\epsilon_1$ over a wide energy
range, essentially as a multi\verb -  (rather than single\verb - )
channel information storage system. The high energy
magnetodielectric (HEMD) effect was recently discovered in several
materials including inhomogeneously mixed-valent K$_2$V$_3$O$_8$
($\sim$5\% at 30 T near 1.2 eV), Kagom\'e staircase compound
Ni$_3$V$_2$O$_8$ ($\sim$16\% at 30 T near 1.3 eV), and hexagonal
multiferroic HoMnO$_3$ ($\sim$8\% at 20 T near 1.8
eV).\cite{Rai2006a, Rai2006b, Rai2006c} In these materials, the HEMD
effect derives from spin-lattice-charge mixing effects. Part of our
continuing strategy to increase  dielectric contrast centers on the
exploitation of electronic mechanisms such  as metal-insulator
transitions, charge ordering, and orbital ordering that are known to
drive changes in the optical constants. \cite{Padilla035120,
Ruckamp2005} Metal-insulator transitions are common with decreasing
temperature, but they are rarely induced by magnetic field. Here, we
report a colossal HEMD response in
(La$_{0.4}$Pr$_{0.6}$)$_{1.2}$Sr$_{1.8}$Mn$_2$O$_7$, behavior that
results from a magnetic field driven spin-glass insulator to
ferromagnetic metal transition. Remnants of the transition also
drive the HEMD effect at room temperature.

\begin{figure}[t]
\vskip -0.5cm
\includegraphics[width = 3.8in]{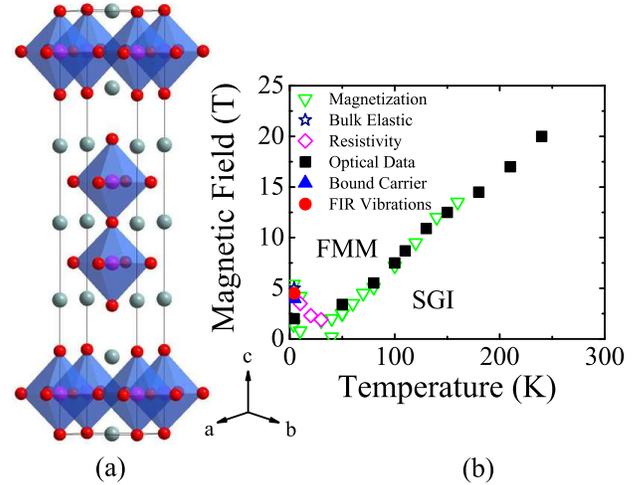}% Here is how to import EPS art
\caption{\label{fig_CMRStru} (Color online) (a) 300 K crystal
structure of (La$_{0.4}$Pr$_{0.6}$)$_{1.2}$Sr$_{1.8}$Mn$_2$O$_7$,
with O (red/black), Mn-containing octahedra (magenta/gray), and
rare/alkaline earth ions (olive/light gray).\cite{Apostu2001} (b)
$H-T$ phase diagram, extracted from optical properties data
($\blacksquare$) for $H\parallel c$. Data are taken with increasing
field. Selected magnetization, magnetostriction, and resistivity
results are shown for comparison. \cite{Cao2006, Gordon2001,
Matsukawa2004, Nakanishi} The spin-glass insulator to ferromagnetic
metal transition is robust at low temperature but diffuse at high
temperatures.}
\end{figure}

(La$_{0.4}$Pr$_{0.6}$)$_{1.2}$Sr$_{1.8}$Mn$_2$O$_7$ derives from the
La$_{1.2}$Sr$_{1.8}$Mn$_2$O$_7$ parent compound, a double-layer
perovskite crystallizing in a body-centered tetragonal structure
(space group $I4/mmm$), as shown in Fig.~\ref{fig_CMRStru}
(a).\cite{Apostu2001} The
(La$_{1-z}$Pr$_{z}$)$_{1.2}$Sr$_{1.8}$Mn$_2$O$_7$ series provides an
opportunity to investigate the physical properties of bilayer
manganites as a function of chemical composition, tuning that gives
rise to a rich phase diagram in this and related
systems.\cite{Tomioka2004, Cao2006} The parent  compound ($z$=0)
displays a paramagnetic insulator to ferromagnetic metal transition
at $T_c$=120 K. With increasing Pr substitution, $T_c$ decreases,
and eventually  disappears ($z$=0.6), establishing a new spin glass
insulating ground state. \cite{Cao2006, Tomioka2004} The
ferromagnetic metallic state that is suppressed by chemical pressure
in (La$_{0.4}$Pr$_{0.6}$)$_{1.2}$Sr$_{1.8}$Mn$_2$O$_7$ is recovered
under magnetic field, as revealed in the $H-T$ phase diagram
(Fig.~\ref{fig_CMRStru} (b)). Transport, magnetization, elastic, and
optical measurements demonstrate that the critical field is $\sim$5
T at 4.2 K, above which the system is in the ferromagnetic metallic
state. \cite{Cao2006, Gordon2001, Matsukawa2004, Nakanishi} The
local structure  changes dramatically with applied  magnetic field.
\cite{Gukasov2005} Both  Mn-O stretching and bending modes soften
through the field-driven transition.\cite{Cao2006}

\begin{figure}[tbh]
\vskip -0.7cm
\includegraphics[width = 3.5 in]{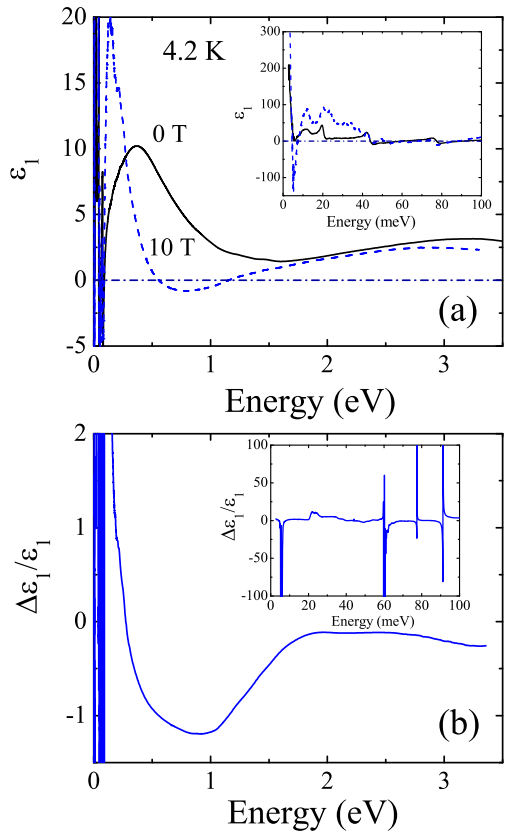}% Here is how to import EPS art
\caption{\label{fig_CMRLTDiel} (Color online) (a) Dielectric
response of (La$_{0.4}$Pr$_{0.6}$)$_{1.2}$Sr$_{1.8}$Mn$_2$O$_7$ for
$H$=0 (solid line) and 10 T (dashed line) ($H\parallel c$) at 4.2 K,
as determined by Kramers-Kronig analysis of the measured reflectance
. (b) The high-energy dielectric contrast, $\Delta
\epsilon_1/\epsilon_1$, for $H$=0 and 10 T.}
\end{figure}

Figure~\ref{fig_CMRLTDiel} (a) displays the real part of the
dielectric response as a function of energy at low temperature. At
zero field, the dispersive response exhibits typical behavior of a
semiconductor, with an overall positive $\epsilon_1$ except for the
regions with strong phonon dispersions. The dispersive response of
the Pr-substituted compound in the high field state is different due
to the change in the ground state. $\epsilon_1$  crosses zero at the
screened plasma energy $\sim$1.2 eV, recrosses into positive
territory at $\sim$0.6 eV, and crosses zero several times again at
low energy due to the dispersive phonon behavior (inset,
Fig.~\ref{fig_CMRLTDiel} (a)). To emphasize the difference between
the zero and high field dielectric response of
(La$_{0.4}$Pr$_{0.6}$)$_{1.2}$Sr$_{1.8}$Mn$_2$O$_7$, we plot the
 dielectric contrast $\Delta
\epsilon_1/\epsilon_1 =[\epsilon_1(E, H)-\epsilon_1(E,
0)]/\epsilon_1(E, 0)$ (Fig.~\ref{fig_CMRLTDiel} (b)). The size of
the dielectric contrast depends on energy. It is as large as
$\sim$100\% near 0.8 eV. This is the range of Mn charge transfer and
on-site excitations. In selected phonon regimes (near 6, 60, 77, 90
meV), the dielectric contrast is enormous, on the order of 10000\%.
The 4.2 K dielectric contrast is a direct consequence of the robust
low temperature spin-glass insulator to ferromagnetic metal
transition and strong spin-lattice-charge cross coupling in
(La$_{0.4}$Pr$_{0.6}$)$_{1.2}$Sr$_{1.8}$Mn$_2$O$_7$.

\begin{figure}[tbh]
\vskip -0.7cm
\includegraphics[width = 3.5 in]{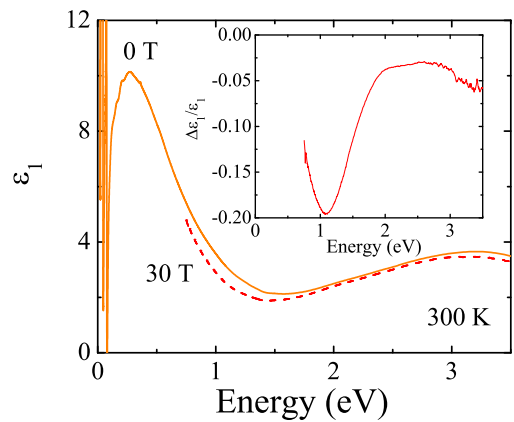}% Here is how to import EPS art
\caption{\label{fig_CMRRTDiel} (Color online) Dielectric response of
(La$_{0.4}$Pr$_{0.6}$)$_{1.2}$Sr$_{1.8}$Mn$_2$O$_7$ for $H$=0 (solid
line) and 30 T (dashed line) ($H\parallel c$) at 300 K, as
determined by Kramers-Kronig analysis of the measured reflectance.
The inset displays the high-energy dielectric contrast, $\Delta
\epsilon_1/\epsilon_1$, for $H$=0 and 30 T. }
\end{figure}

It is clearly desirable to design and assemble multifunctional
devices that operate at higher temperatures. Materials that exhibit
room temperature magnetoelectric coupling are therefore needed. To
this end, we have employed the remnants of the spin-glass insulator
to ferromagnetic metal transition in
(La$_{0.4}$Pr$_{0.6}$)$_{1.2}$Sr$_{1.8}$Mn$_2$O$_7$ to achieve a 300
K HEMD effect. Figure~\ref{fig_CMRRTDiel} displays the room
temperature dielectric properties of
(La$_{0.4}$Pr$_{0.6}$)$_{1.2}$Sr$_{1.8}$Mn$_2$O$_7$ at 0 and 30 T.
The dispersive response shifts to lower energy in applied magnetic
field, consistent with the previous observation of redshifted
oscillator strength in the ferromagnetic metallic
state.\cite{Cao2006} Strikingly, the dielectric contrast
%, $\Delta\epsilon_1/\epsilon_1$,
displays a 20\% change near 1.1 eV (inset of
Fig.~\ref{fig_CMRRTDiel}).\cite{RTDielLowEnergy} Although weaker
than that at 4.2 K, this finding demonstrates that electronic
transitions can be harnessed for 300 K HEMD effects.

The HEMD effect in
(La$_{0.4}$Pr$_{0.6}$)$_{1.2}$Sr$_{1.8}$Mn$_2$O$_7$ is substantially
larger than that in  mixed-valent K$_2$V$_3$O$_8$, Kagom\'e
staircase compound Ni$_3$V$_2$O$_8$, and frustrated HoMnO$_3$,
oxides where magnetoelastic/magnetoelectric coupling is
important.\cite{Rai2006a, Rai2006b, Rai2006c} As mentioned
previously, this stronger dielectric contrast is directly related to
exploiting the field-induced insulator to metal transition.
Reviewing the Tomioka-Tokura global phase diagram in complex
perovskites, \cite{Tomioka2004} the title compound is sitting in
close proximity to the phase boundary and is therefore susceptible
to physical tuning - even when the transition itself is no longer
well-defined. Control of disorder therefore  provides an important
route for tuning the HEMD effect. \cite{Cao2006, Tomioka2004}

\begin{figure}[tbh]
\vskip -0.7cm
\includegraphics[width = 3.5 in]{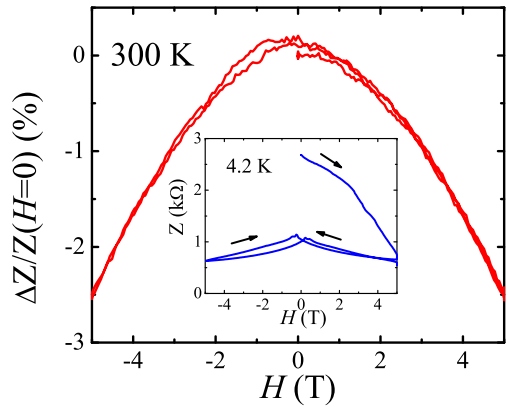}% Here is how to import EPS art
\caption{\label{fig_CMRStatic} (Color online) The magneto-impedance
contrast of (La$_{0.4}$Pr$_{0.6}$)$_{1.2}$Sr$_{1.8}$Mn$_2$O$_7$ as a
function of magnetic field  ($H\parallel c$) at 300 K. The inset
displays the magnitude of the impedance  as a function of magnetic
field ($H\parallel c$) at 4.2 K. The sample was zero-field cooled. }
\end{figure}

It is interesting to compare the HEMD effect in
(La$_{0.4}$Pr$_{0.6}$)$_{1.2}$Sr$_{1.8}$Mn$_2$O$_7$ with static
results.  Due to the large DC conductivity of these single crystal
samples, we investigate the low frequency magnetic response in terms
of the impedance rather than $\epsilon_1$. \cite{Qin2002,
Castro2004} The inset of Fig.~\ref{fig_CMRStatic} displays the low
frequency (30 kHz) impedance as a function of magnetic field at 4.2
K. The high impedance insulating state at 0 T is suppressed with
increasing magnetic field, leading to a large change in impedance
between 0 and 5 T.  As the field decreases towards 0 T, there is a
small increase in impedance, but the sample remains trapped in the
metastable low impedance state. The phase of the complex impedance
displays similar behavior.  Field-induced transitions and trapping
in a metastable state have been observed in resistivity measurements
of (La$_{0.4}$Pr$_{0.6}$)$_{1.2}$Sr$_{1.8}$Mn$_2$O$_7$,
\cite{Matsukawa2005} and were attributed to the strong
magnetoresistive coupling in this system. This static spin-charge
coupling persists to higher temperatures. The main panel of
Fig.~\ref{fig_CMRStatic} displays the 300 K magneto-impedance
contrast as a function of magnetic field. The impedance of the
sample changes up to 2.5\% in a 5 T applied field. This
magneto-impedance is completely non-hysteretic, different from that
at low temperatures. Similar magneto-impedance has been observed
close to the ferromagnetic ordering temperature in other manganites,
\cite{Qin2002, Castro2004} and was attributed to the dependence of
the skin depth on both magnetic field and frequency. Clearly, these
room temperature effects represent significant spin-charge coupling
persisting to temperatures well above the magnetic transition
temperature, in line with the 300 K HEMD effect.

To summarize, we observed a large HEMD effect in the bilayer
manganite (La$_{0.4}$Pr$_{0.6}$)$_{1.2}$Sr$_{1.8}$Mn$_2$O$_7$,  a
direct consequence of field driven spin-glass insulator to
ferromagnetic metal transition. The remnants of the transition can
be used to achieve dielectric contrast at room temperature. This
discovery suggests that  electronic mechanisms such as the
metal-insulator transition, charge ordering, and orbital ordering
can be exploited to give substantial dielectric contrast in other
materials.

\vspace{1cm}

\emph{\large{Experimental}}
\\

\footnotesize{}

Single crystals of
(La$_{0.4}$Pr$_{0.6}$)$_{1.2}$Sr$_{1.8}$Mn$_2$O$_7$ were grown from
sintered rods of same nominal composition by the floating-zone
technique, using a mirror furnace.\cite{Apostu2001} Typical crystal
dimensions were $\approx$4$\times$5$\times$2 mm$^3$. They were
cleaved to yield a shiny surface corresponding to the $ab$ plane.

Near normal $ab$ plane  reflectance of
(La$_{0.4}$Pr$_{0.6}$)$_{1.2}$Sr$_{1.8}$Mn$_2$O$_7$ was measured
over a wide energy range (3.7 meV - 6.5 eV) using a series of
different spectrometers as described previously.\cite{Cao2006} %The
%spectral resolution was 2 cm$^{-1}$ in the far and middle-infrared
%and 2 nm in the near-infrared, visible, and near-ultraviolet. Low
%temperature spectroscopies were carried out with a continuous-flow
%helium cryostat and temperature controller.
A Kramers-Kronig analysis was employed to obtain the optical
constants,\cite{Wooten1972} particularly the dispersive part of the
dielectric response  as ${\tilde{\epsilon}}(E) =
{\epsilon}_1(E)+i{\epsilon_2}(E)$. The magneto-optical properties
were measured at the National High Magnetic Field Laboratory in
Tallahassee, FL, using two different spectrometers and both
superconducting and resistive magnets.\cite{Cao2006} Experiments
were performed at 4.2  and 300 K for $H\parallel c$. \cite{Cao2006}
The field-induced changes in the measured reflectance were studied
by taking the ratio of reflectance at each field and reflectance at
zero field, i.e., [R($H$)/R($H$=0 T)]. To obtain the high field
dielectric properties, we renormalized the zero-field absolute
reflectance with the high-field reflectance ratios, and recalculated
$\epsilon$$_1(E)$ using Kramers-Kronig techniques.\cite{Wooten1972}
%To facilitate comparison with static magnetodielectric measurements,
%we define the magnetodielectric contrast as $\Delta
%\epsilon_1/\epsilon_1 =[\epsilon_1(E, H)-\epsilon_1(E,
%0)]/\epsilon_1(E, 0)$.

Impedance measurements were carried out at 30 kHz, using an Agilent
4284A LCR meter in a two-wire configuration with the AC current
parallel to the applied magnetic field. The high temperature
impedance was corrected for the contribution from the probe, which
is negligible at 4.2 K.

\vspace{1cm}

\emph{\large{Acknowledgements}}
\\

Work at the University of Tennessee is supported by the Materials
Science Division, Basic Energy Sciences, U.S. Department of Energy
(DE-FG02-01ER45885). %Work at Wayne State University is supported by
%the Jane and Frank Warchol Foundation.
The international aspects of this research were supported by the
National Science Foundation (INT-019650) at UT and the Jane and
Frank Warchol Foundation at Wayne State. A portion of this work was
performed at the NHMFL, which is supported by NSF Cooperation
Agreement DMR-0084173 and by the State of Florida.


\begin{thebibliography}{JinboCao}
\bibitem{RTDielLowEnergy}
The HEMD effect in the far and middle infrared region could not be
assessed in our experiments at the NHMFL. We anticipate that the
dielectric contrast will be larger based upon its low temperature
behavior.
\end{thebibliography}
\end{document}